\documentclass[useAMS,usenatbib]{mnras}
\usepackage{graphicx,lscape,amssymb}
\usepackage{natbib}
\usepackage{fancyhdr}
\usepackage{subfigure}
\usepackage{amsmath}
\usepackage{mathtools}
\usepackage{tabulary}
\usepackage{color}
\usepackage{booktabs, dcolumn}

\newcommand{\Teff}{$T_{\mathrm{eff}}\,$}
\newcommand{\Msun}{$M_{\mathrm{\odot}}\,$}

\title[Convection in magnetic white dwarfs] {Can magnetic fields suppress convection in the atmosphere of cool white dwarfs? A case study on WD2105$-$820}

\author[Gentile Fusillo et al.]{N. P. Gentile Fusillo$^1$, P.-E. Tremblay$^1$,  S. Jordan$^2$, B. T. G\"ansicke$^1$,
\newauthor J. S. Kalirai$^{3,4}$, J. Cummings$^4$\\
$^1$Department of Physics, University of Warwick, Coventry, CV4 7AL, UK\\
$^2$Astronomisches Rechen-Institut, Zentrum f\"ur Astronomie der Universit\"at Heidelberg, D-69120 Heidelberg, Germany\\
$^3$Space Telescope Science Institute, 3700 San Martin Drive, Baltimore, MD 21218, USA\\
$^4$Center for Astrophysical Sciences, Johns Hopkins University, 3400 N Charles Street, Baltimore, MD 21218, USA}

\begin{document}
\maketitle
\label{firstpage}

\begin{abstract}
Around 10\% of white dwarfs exhibit global magnetic structures with fields ranging from 1 kG to hundreds of MG. Recently, the first radiation magnetohydrodynamics simulations of the atmosphere of white dwarfs showed that convection should be suppressed in their photospheres for magnetic fields with strengths $B \gtrsim 50$ kG. These predictions are in agreement with our knowledge of stellar physics (e.g. energy transfer in strong magnetic field regions of the solar photosphere), but have yet to be directly confirmed from white dwarf observations. We obtained COS far-UV spectroscopy of the weakly magnetic, hydrogen-atmosphere, white dwarf WD2105$-$820 and of three additional non-magnetic, convective remnants (all in the \Teff range 9000$-$11\,000 K). We fitted both the COS and the already available optical spectra with convective and radiative atmospheric models. As expected, we find that for two of the non-magnetic comparison stars only convective model fits predict consistent \Teff values from both the optical and the FUV spectra. In contrast, for  WD2105$-$820 only the best fitting radiative model produced consistent results.
\end{abstract}


\section{Introduction}
Nearly half a century ago, \citet{kempetal70-1} detected circularly polarized light from  the white dwarf GJ\,742 proving that the star harbored a magnetic field. Since then hundreds of magnetic white dwarfs have been identified via spectropolarimetry (e.g. \citealt{friedrichetal96-3, vennesetal03-1, kawkaetal07-1}) or detection of Zeeman splitting (e.g. \citealt{hagenetal87-1, reimersetal96-1, gaensickeetal02-5, Kleinmanetal13-1}). The actual fraction of white dwarfs with magnetic fields remains extremely hard to constrain with estimates  of 3$-$4 per cent in a magnitude limited census \citep{liebertetal03-1, kepleretal13-1} and 10 to 30 per cent \citep{kawkaetal07-1} for the local volume limited sample. Recent work on metal polluted white dwarfs with \Teff$<8000$ has reported a magnetic incidence of $13\pm 4$ per cent, much higher than among hot white dwarfs. Consequently the striking higher magnetic incidence observed in the solar neighborhood may be caused by the fact that the local sample is dominated by cool white dwarfs.
The majority of these objects have magnetic fields with strengths $B>1$\,MG \citep{schmidtetal03-1, kulebietal09-1}. These high-field  white dwarfs exhibit obvious Zeeman line splitting and are easily identified in large area spectroscopic surveys (e.g. SDSS; \citealt{kepleretal13-1}). These split spectral line profiles are unsuitable for the standard spectroscopic technique
employed to derive atmospheric parameters from the Balmer
lines \citep{bergeronetal92-1} and, as a result, the masses and cooling ages of most magnetic white dwarfs are only weakly constrained. However, spectrum independent measurements of mass for high-field magnetic white dwarfs can be derived for objects with measured trigonometric parallaxes or in wide binaries. Recent studies have shown these stars to be more massive than their non-magnetic counterparts (mean mass of $\simeq 0.8$\,\Msun in contrast with $\simeq0.6$\,\Msun for non-magnetic white dwarfs), however, the number of magnetic white dwarfs with reliable mass determinations is still very small     \citep{ferrarioetal15-1}.
Weaker magnetic fields ($B\lesssim1$\,MG) are also found in white dwarfs. Even with high-resolution spectroscopy, Zeeman splitting becomes undetectable for fields $B\lesssim20$\,kG \citep{jordanetal07-1} and spectrapolarimetry becomes the only method to identify these magnetic white dwarfs. As a result the actual incidence of low-field magnetic white dwarfs is uncertain, but a few small spectropolarimetric surveys currently put the fraction of 1-100\,kG magnetic white dwarfs at 3-30 per cent \citep{jordanetal07-1, kawka+vennes12-1, landstreetetal12-1}.
Unlike their high-field counterparts, low-field magnetic white dwarfs appear to have an average mass close to that of non magnetic white dwarfs \citep{jordanetal07-1}.

To date the origin of magnetic white dwarfs remains an open question with the main formation scenarios being: fossil fields from magnetic peculiar Ap and Bp stars \citep{angel81-1,wickramasinghe+ferrario00-1}, the result of the merger of two white dwarfs \citep{kulebietal13-1, wickramasingheetal14-1}, or the product of a  magnetic dynamo generated within the common envelope during the evolution of a binary system \citep{toutetal08-1}, or in main sequence stars with convective cores \citep{stelloetal16-1, cantielloetal16-1}.

The second data release of Gaia (DR2) will include precise stellar parallaxes for all known magnetic white dwarfs \citep{torresetal05-1, carrascoetal14-1}, and will allow us to identify $\simeq 400\,000$ new white dwarfs, among which there will be thousands of magnetic systems.
Thanks to Gaia it will be possible to measure with unprecedented accuracy the mass distribution of white dwarfs, and possibly tackle many unanswered questions regarding the incidence, mass distribution and origin of magnetic remnants. 
Magnetic white dwarfs significantly contribute to the global white dwarf population, particularly in the high-mass regime. Therefore, in order to correctly derive the Galactic stellar formation history and the initial mass function, it is of paramount importance to be able to fully characterize these magnetic objects.

In agreement with our general knowledge of stellar physics (e.g. sunspots), many authors have suggested that convection is completely inhibited in high-field magnetic white dwarfs \citep{wickramasinghe+martin86-1, valyavinetal14-1}. Recently \citet{tremblayetal15-1} performed the first radiation magnetohydrodynamics simulations of the atmosphere of white dwarfs,	 which confirmed that convection should be suppressed in the photospheres of these objects for magnetic fields with strengths $B \gtrsim 50$\,kG \citep{tremblayetal15-1}. However, despite the robust theoretical evidence, these predictions have not yet been confirmed by direct observations. 

In this work we present the case study of the weakly magnetic white dwarf WD2105$-$820 \citep{landstreetetal12-1}. We compare the ultraviolet (UV) and optical spectrum of this star, as well as the available optical and near infra-red photometry with both convective and radiative atmospheric models and assess which one provides the most consistent description across all wavelengths. For comparison we also carry out the same analysis on three additional non-magnetic, convective white dwarfs. 

\section{Targets}
Hydrogen atmosphere white dwarfs become convective for \Teff$\lesssim 14\,000$K at log\,$g = 8$ \citep{tremblayetal13-1}, as the recombination of hydrogen causes a significant increase in the opacity. However, above \Teff$\simeq12\,000$\,K convection is not yet fully dominant, and convective models, as well as radiative ones, where convection was artificially suppressed, are very similar, particularly in the UV. On the other hand below \Teff$\simeq9000$\,K white dwarfs have very small UV fluxes and, as we show later in sect.\,\ref{con_vs_rad}, UV observations are key for distinguishing between convective and radiative atmospheres.

At field strengths, $B$, above 100\,kG, Zeeman broadening significantly distorts the spectrum of a white dwarf and an accurate treatment of the combined Stark and Zeeman broadening becomes necessary to obtain stellar parameters from optical spectroscopy, even at medium resolution ($\sim3 \mathrm{\AA}$). Such model does not currently exists and as a result atmospheric parameters of high-field magnetic white dwarfs are often very uncertain. Furthermore, in this high-field regime ($B>$ 1\,MG), because of the splitting of the Lyman alpha line and of the quasi-molecular satellites \citep{allard+kielkopf09-1}, the interpretation of far-UV (FUV) observations would also require a dedicated magnetic analysis. Finally, even assuming a radiative atmosphere, magnetic pressure and magneto-optical effects may become important with stronger fields and a simple radiative model (where convective flux is fixed to zero, see sect.\,\ref{model_sect}) may not correctly describe the atmospheric structure. 

In conclusion, in order to observationally test whether magnetic fields can suppress convection in white dwarf atmospheres, we need to identify suitable magnetic and non magnetic white dwarfs within a narrow range of temperatures and magnetic field strengths. 
In this initial selection we picked the white dwarf WD2105$-$820 (\Teff$ = 10\,389$\,K and log\,$g=8.01$; Table\,\ref{all_teff}). \citet{koesteretal98-1} first noticed excess broadening in the core of H$\alpha$ in high-resolution UVES spectroscopy (0.26\,\AA) of this star, which they suggested could be caused by a magnetic field with average strength $B\simeq43$\,kG. 
Over a decade later \citet{landstreetetal12-1} observed WD2105$-$820 as part of their spectropolarimetric survey of cool white dwarfs and measured a constant longitudinal magnetic field $B_{z}\simeq 9.5$\,kG, confirming the magnetic nature of the star. The $B_{z} / |B|$ ratio of $\simeq0.22$ indicates a dipolar morphology \citep{landstreet88-1, schmidt+norsworthy91-1} and \citet{landstreetetal12-1} concluded that the most likely magnetic field structure for WD2105$-$820 is  a simple centered dipole with
a polar field strength of $\simeq56$\,kG, and magnetic axis parallel to
the rotation axis inclined at $\simeq68$\,deg with respect to the line
of sight.

The average field strength of $\simeq43$\,kG is too small to produce any visible Zeeman splitting in medium resolution spectroscopy; consequently the standard spectroscopic method to evaluate atmospheric parameter by comparing the Balmer line profiles with model atmospheres can still be reliably used for WD2105$-$820. For the comparison we also selected three additional cool, apparently non-magnetic, single, DA white dwarfs: WD0839$-$327, WD1544$-$374, WD1310+583.

\section{Observations}
We obtained FUV spectroscopy with	the Cosmic Origins Spectrograph (COS, \citealt{greenetal12-1}) on the Hubble Space Telescope, using the G140L grating (central wavelength setting $1105\, \mathrm{\AA}$) for our four chosen stars. This set up only uses segment A of the detector and covers the wavelength range from $1122\, \mathrm{\AA}$ to $2148\, \mathrm{\AA}$. WD2105$-$820, WD1544$-$374, and  WD1310+583 where observed between  March 5th and May 3rd 2016, with exposure time between 2152 and 2460\,s as part of proposal ID 14214 PI: P.-E. Tremblay.  WD0839$-$327 was instead observed as part of proposal 14076 PI: B. G\"ansicke on Jan 30, 2016. For our analysis we used the spectra reduced using the COS calibration pipeline ({\sc{calcos}}) provided by the  Mikulski Archive for
Space Telescopes (MAST).

\section{The models}
\label{model_sect}
The four stars we consider in this study were all included in \citet{gianninasetal11-1} accurate analysis of bright hydrogen atmosphere white dwarfs. \citet{gianninasetal11-1} obtained atmospheric parameters by fitting medium resolution spectra (3-6\,\AA) of these white dwarfs with 1D pure hydrogen, plane-parallel model atmospheres where energy transport by convection is included following the ML2/$\alpha$ = 0.8 prescription \citep{tremblayetal11-1}. These models also include the improved Stark broadening profiles from \citet{tremblayetal09-1}. 
To account for the known inaccuracies in the 1D mixing-length approach we also apply the 3D corrections developed by \citet{tremblayetal13-1} to the atmospheric parameters published by \citet{gianninasetal11-1} (Table\,\ref{all_teff}).
In order to assert the effect of magnetic fields on convection in white dwarf atmospheres we also computed a separate grid of purely radiative hydrogen model atmospheres for which we have enforced a convective flux of zero when solving for the atmospheric stratification. We note that for these models 3D corrections are not relevant.

\begin{figure}
\centering 
\includegraphics[width=0.85\columnwidth]{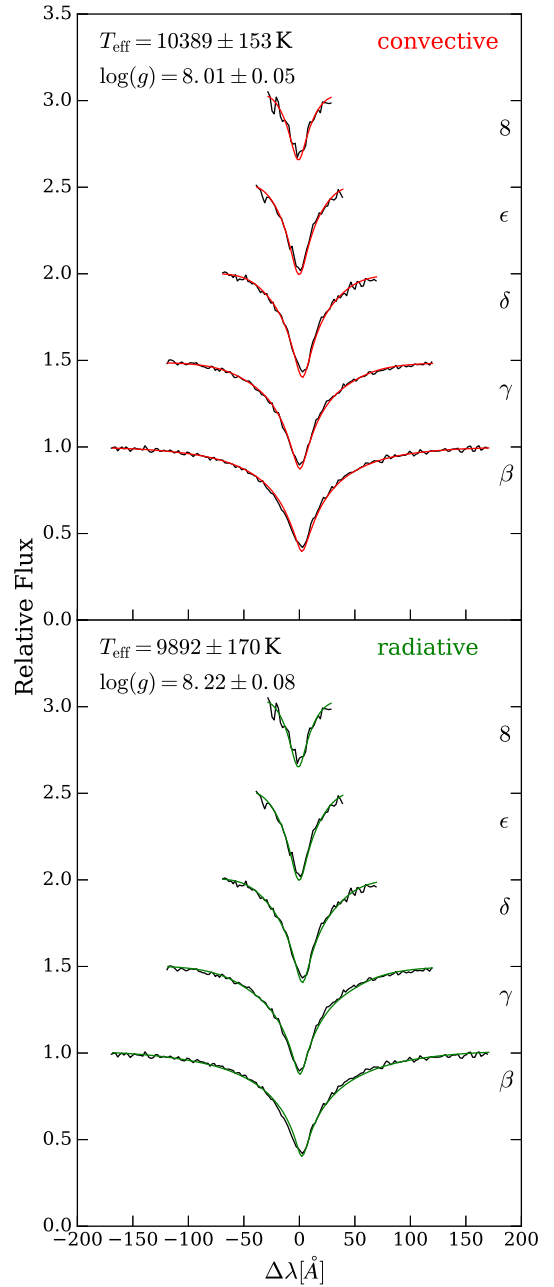}
\caption{\label{rad_fit} Best model fits using a convective  (red, top panel) and a radiative model (green, bottom panel) to the individual Balmer line profiles (black) of the magnetic white dwarf WD2105$-$820. The normalized lines are vertically offset for clarity. The optical spectrum available does not cover the H$\alpha$ line. The best-fit values of \Teff and log\,$g$ are indicated at the top of each panel. Both best-fitting models correctly reproduce the Balmer line profiles, but at significantly different \Teff values.}
\end{figure}

\begin{table*}
\centering
\caption{\label{all_teff} \Teff values obtained from both convective and radiative model atmosphere fits of the optical spectra and COS FUV spectra of the four white dwarfs analysed in this article. Convective \Teff values from optical fits are those published by \citet{gianninasetal11-1}  with additional 3D corrections \citep{tremblayetal13-1}. In the convective FUV fits we directly used $\langle3D\rangle$ spectra \citep{tremblayetal13-1}.}
\begin{tabular}{ll D{?}{\,\pm\,}{5.3} D{?}{\,\pm\,}{5.3} D{?}{\,\pm\,}{5.3} D{?}{\,\pm\,}{5.3}}
\hline
& & \multicolumn{2}{c}{\textit{convective}}& \multicolumn{2}{c}{\textit{radiative}}\\
\hline
Name & & \multicolumn{1}{c}{\textbf{optical \Teff [K]}} & \multicolumn{1}{c}{\textbf{FUV \Teff [K]}}& \multicolumn{1}{c}{\textbf{optical \Teff [K]}} & \multicolumn{1}{c}{\textbf{FUV \Teff [K]}}\\
\hline \\[-1.5ex]
\multicolumn{2}{l }{WD2105$-$820 (magnetic)} & 10\,389?153 & 9812?27& 9892?170 & 10\,047?28\\
\multicolumn{2}{l }{WD0839$-$327} & 9128?132 & 9303?35 & 8676?140 & 9606?42\\ 
\multicolumn{2}{l }{WD1544$-$377} & 10\,394?150 & 10\,598?31 & 9770?161 & 11\,052?37\\

\multicolumn{2}{l }{WD1310+583} & 10\,479?160 & 11\,656?24 & 9728?98 & 12\,042?35\\
\hline\\[-1.5ex]

\end{tabular}
\end{table*}

\section{Convective or radiative}
\label{con_vs_rad}
\begin{figure*}
\hspace{1.cm}\includegraphics[width=0.97\columnwidth]{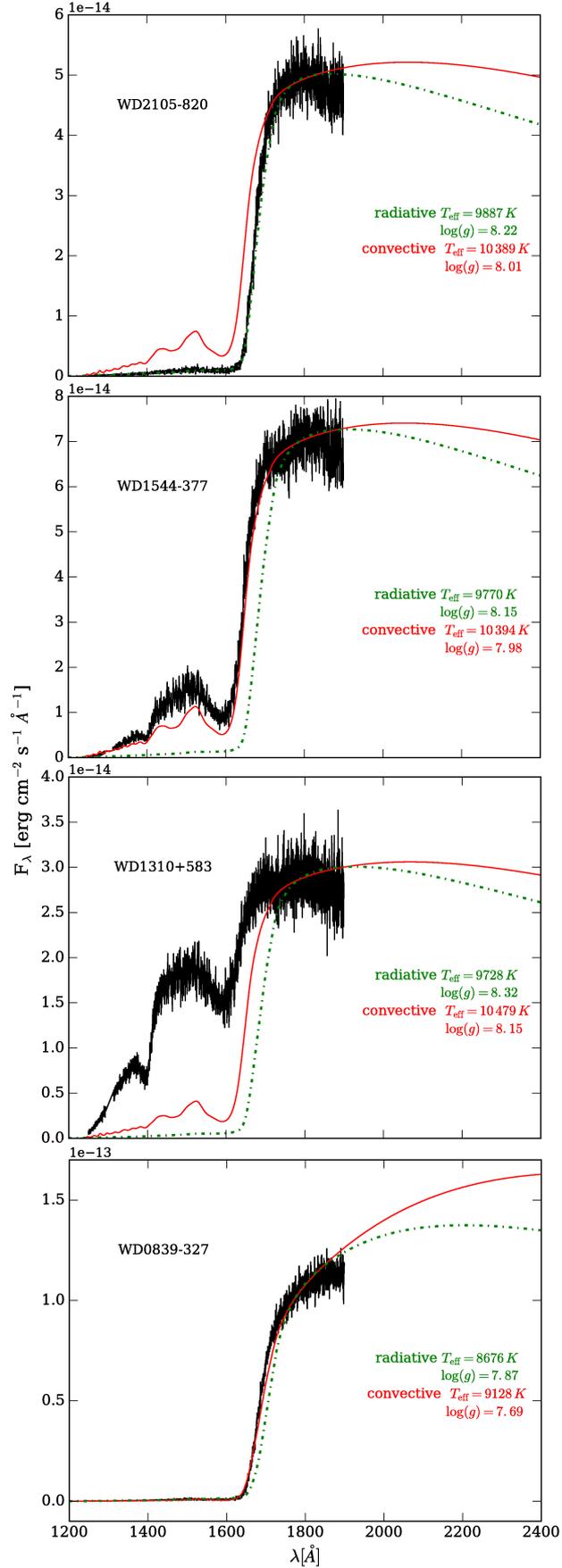}\caption{\label{COS_fit}Far UV COS spectra of WD2105$-$820, WD1544$-$377, WD1310+583 and WD0839$-$327 (black). The radiative and convective models which provide the best fit to the \underline{optical spectra}	 (Table \ref{all_teff}) are shown in green (dashed) and red (solid) respectively.}
\end{figure*}

We fitted the \citet{gianninasetal11-1} optical spectra of our four DA white dwarfs using our radiative models and compared our results with those obtained by \citet{gianninasetal11-1}  (plus 3D corrections). 
As shown in Fig.\,\ref{rad_fit} for WD2105$-$820, both the radiative and the convective models can reproduce the Balmer line profiles of the white dwarf spectra equally well, though at significantly different  \Teff values. However, looking at the COS spectrum of WD2105$-$820 (Fig.\,\ref{COS_fit}) it is immediately obvious that while the best fitting optical radiative model (\Teff$ = 9887$\,K and log\,$g=8.22$) successfully reproduces the FUV spectral profile of WD2105$-$820, the corresponding convective model solution (\Teff$ = 10\,389$\,K and log\,$g=8.01$) fails to do so (Fig.\,\ref{COS_fit}). The  FUV model comparison illustrates how COS spectroscopy can reliably be used to differentiate radiative and convective atmospheres and indicates that convection may indeed be suppressed in WD2105$-$820. 

\begin{figure}
\includegraphics[width=\columnwidth]{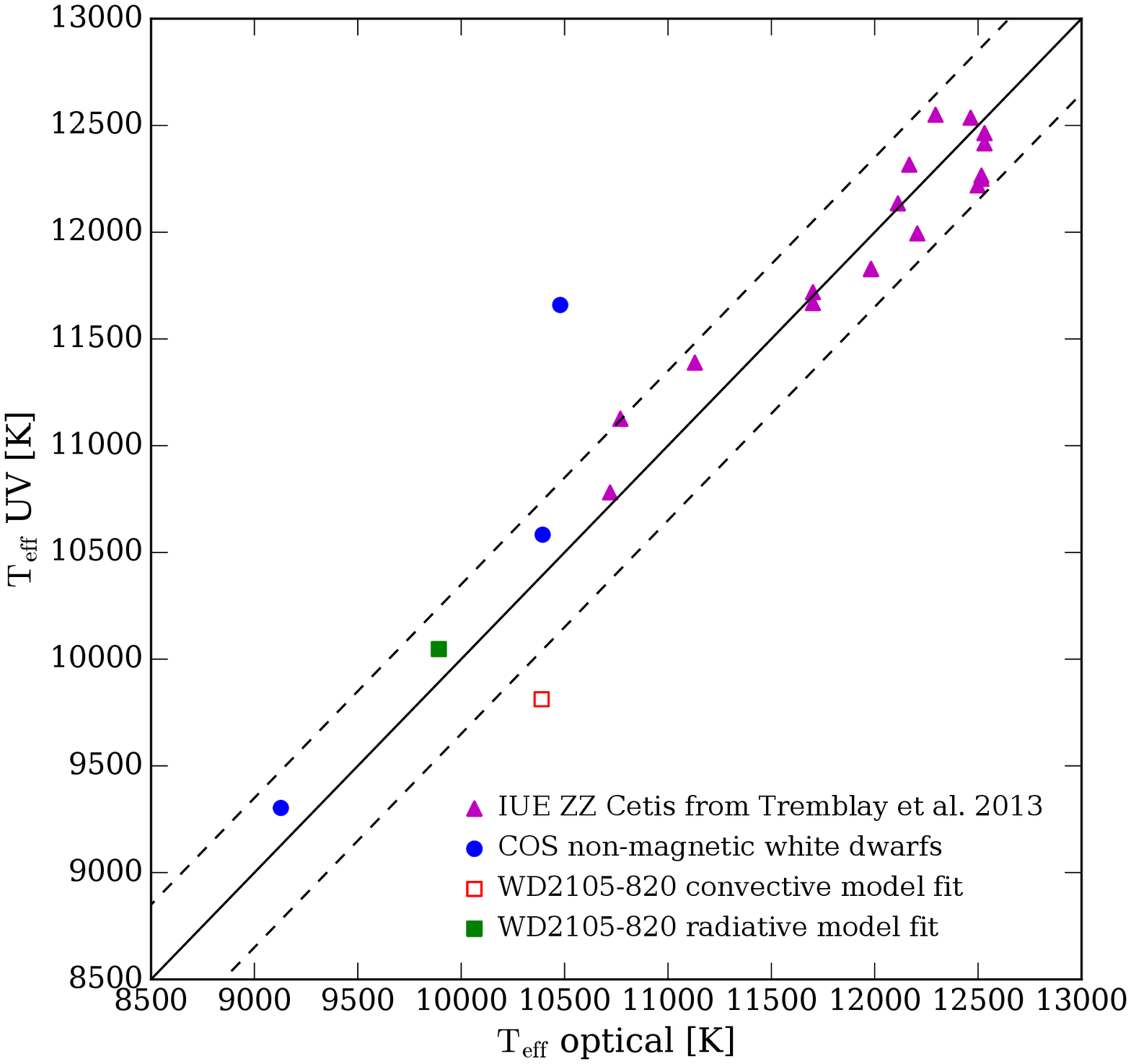}\caption{\label{topVtuv}Comparison of \Teff values derived from UV and optical spectra for the white dwarfs analyzed in this work. Optical log\,$g$ values are assumed in the determination of the UV temperatures and 3D corrections (or $\langle 3D \rangle$ spectra) have been used in all convective model fits. The solid line represents a perfect match between UV and optical temperatures, while the dashed lines represent the $\pm350$\,K uncertainty typically allowed by the optical analysis \citep{tremblayetal13-1}. For comparison we also include the ZZ\,ceti stars analyzed in \citet{tremblayetal13-1} as magenta triangles. The significantly offset blue point represents WD1310+583 (see sect.\ref{double_wd}).}
\end{figure}

A more rigorous way to compare the consistency of optical and FUV data is to fit the COS spectrum with the same atmospheric models (convective and radiative) and compare the \Teff values obtained with the optical ones. Since there are no lines sensitive to gravitational broadening in the FUV spectrum of our stars, in the determination of the FUV temperatures we have to fix the log\,$g$ at the value determined by the optical fit.
As we show in Fig.\,\ref{topVtuv} the best fitting convective models provide consistent \Teff values for the optical and FUV spectra for two of our stars (WD0839$-$327, WD1544$-$374), but significantly different ones ($>3\sigma$) for the remaining two white dwarfs (WD2105$-$820, WD1310+583).
In contrast, the best fit radiative models give a much better agreement for the optical and FUV \Teff values of the magnetic white dwarf WD2105$-$820, but result in a marked disagreement for WD0839$-$327 and WD1544$-$374 (Table\,\ref{all_teff}). This result is in agreement with our prediction and  indicates that, despite being cool enough to have a convective atmosphere (like WD0839$-$327 and WD1544$-$374), the magnetic white dwarf WD2105$-$820 is better characterized by a purely radiative model suggesting that convection in WD2105$-$820 is suppressed by the presence of the magnetic field. 

An open question remains about WD1310+583 for which neither the convective nor the radiative models could provide \Teff. In sect.\,\ref{double_wd} we speculate that this system may in fact be an unresolved double degenerate.

\section{comparison with photometry}
\label{photo_compare}
\begin{figure}
\includegraphics[width=\columnwidth]{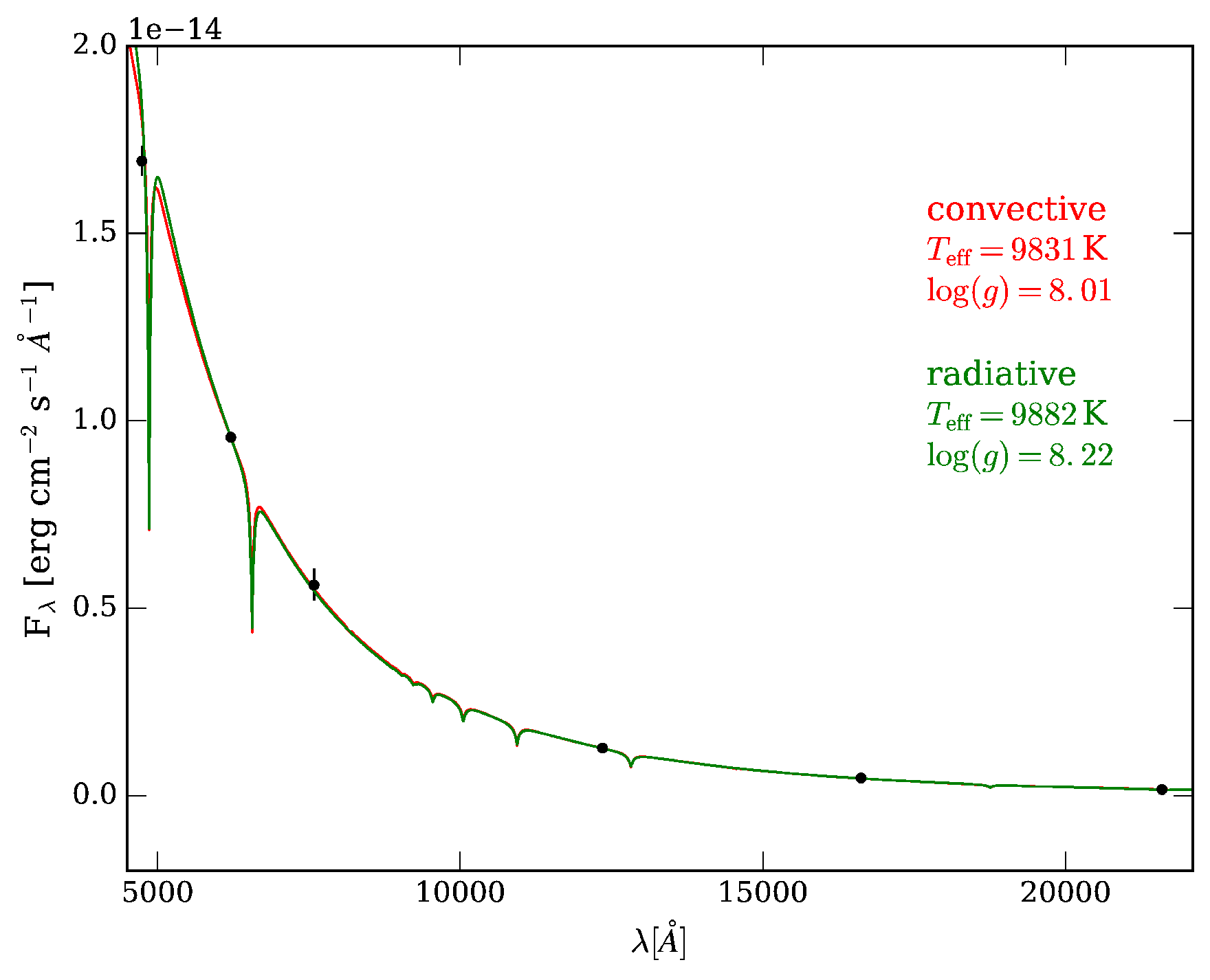}\caption{\label{photofit} Best fits to the $griJHK$ photometry of WD2105$-$820 using convective and radiative models.}
\end{figure}

As an additional test of our spectroscopic analysis of WD2105$-$820 we compared the photometric \Teff of this star with the \Teff obtained from convective and radiative spectroscopic model fits.
We retrieved optical $gri$ band photometry from the AAVSO Photometric All-Sky Survey Data Release 9 (APASS DR9; \citealt{hendenetal15-1}) and near infrared $JHK$ band photometry from the Two Micron All-Sky Survey (2MASS; \citealt{2MASS06-1}).
We then fitted the photometric spectral energy distribution (SED) with both radiative and convective atmospheric models to obtain a photometric \Teff estimate. Photometric model fits are not sensitive to log\,$g$ so, when determining the best fitting model, we fixed the log\,$g$ to the value found by the optical spectroscopic fit. 
As shown in Fig.\,\ref{photofit} the photometric fits are mostly independent of both the type of model used and the adopted log\,$g$ values. Consequently, we obtain very similar \Teff estimates for both the convective and radiative best fits (\Teff$=9831\pm210$\,K and \Teff$=9882\pm190$\,K respectively). In both cases the photometric \Teff is consistent with the spectroscopic radiative solution (both optical and FUV) and it is over 500\,K cooler than the spectroscopic convective solutions.

In conclusion, photometric observations are in agreement with our spectroscopic analysis and corroborate the theory that convection is suppressed in the atmosphere of the magnetic white dwarf WD2105$-$820. 

\section{WD1310+583: a double white dwarf?}
\label{double_wd}
\begin{figure}
\includegraphics[width=\columnwidth]{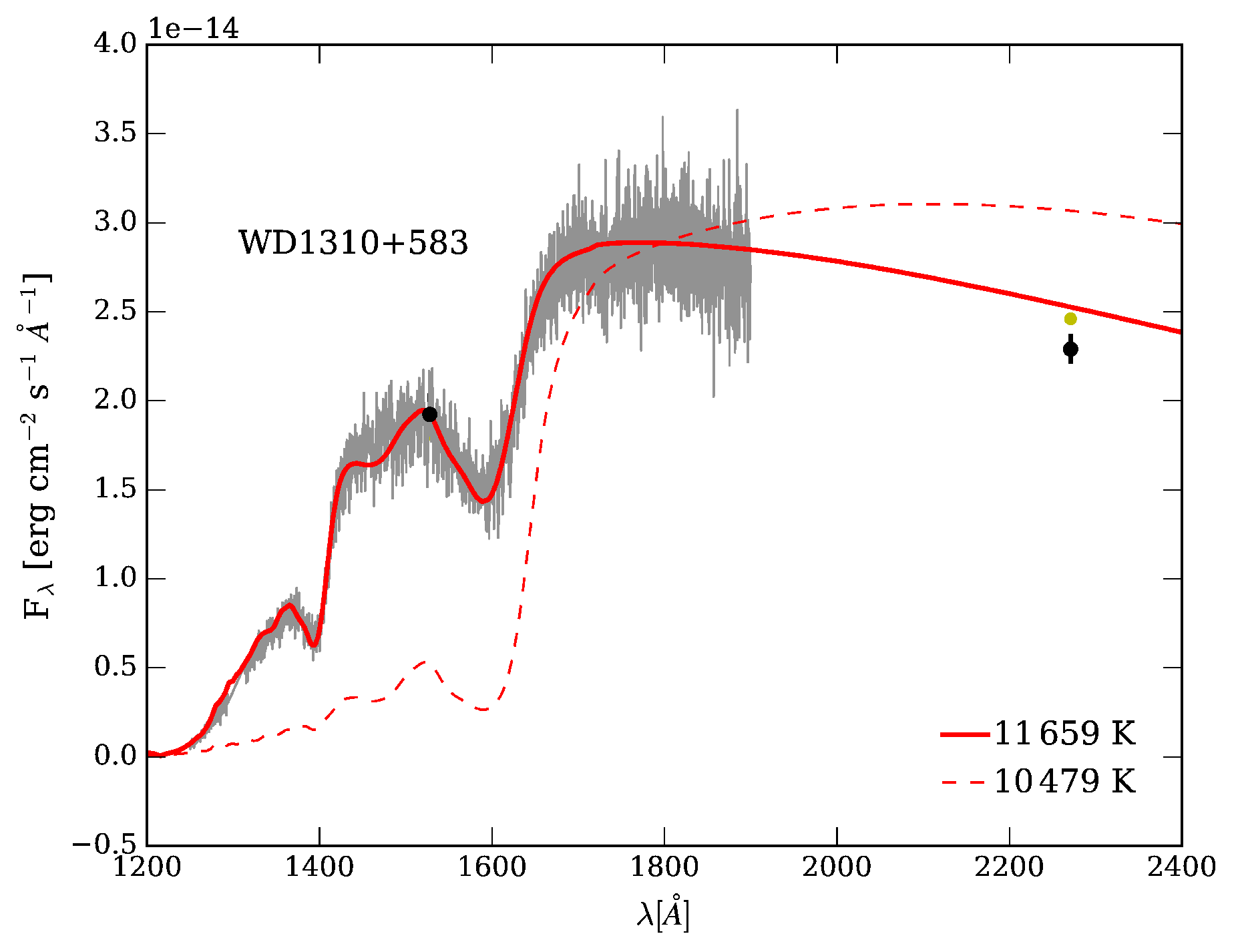}\caption{\label{WD1310_compare} COS spectrum of WD1310+583. Best fitting FUV and optical models are shown respectively in solid and dashed red overlay. Galex $fuv$ and $nuv$ observations are represented by the blue point and synthetic $fuv$ and $nuv$ fluxes calculated from the hotter model are show in yellow.}
\end{figure}

The best convective model fit to the FUV spectrum of WD1310+583 provides a \Teff of 11\,598\,K while the best fit to the optical spectrum indicates a value of only 10\,479\,K (Fig.\,\ref{topVtuv}). Adopting radiative model atmospheres does not improve the comparison and still yields a difference between the optical and FUV \Teff of $\simeq2000$\,K. 
We retrieved $ugriz$ optical photometry from the Sloan Digital Sky Survey (SDSS DR13; \citealt{SDSS_DR1316-1}) and 2MASS $JHK$ NIR photometry for this white dwarf and fitted it with convective model atmospheres. We obtained a photometric \Teff of 10\,250\,K which  confirms the value obtained from the optical spectroscopic fit. 
Similarly, comparing the Galex UV  photometry of WD1310+583 (corrected for non linear detector response;  \citealt{camarotaetal14-1}) with the COS spectrum, we find that the $fuv$ flux perfectly matches the spectrum and, while the $nuv$ flux does not exactly agrees with the $11\,659$\,K model, it seems to rule out the $10\,479$\,K solution (Fig.\,\ref{WD1310_compare}).
It is clear, therefore, that the offset we see in FUV and optical \Teff values is not caused by problems with the spectroscopy and we are truly measuring two different effective temperatures.
We conclude that WD1310+583 is most likely a double degenerate binary system composed of a hotter white dwarf which dominates the FUV emission and a cooler white dwarf whose contribution becomes significant only in the optical. In this scenario all the examined spectra of WD1310+583 are a combination of the ``real" spectra of the two white dwarfs. We can therefore infer that the FUV \Teff we obtained represents a lower limit of the actual \Teff of the hotter white dwarf. 

\begin{figure*}
\includegraphics[width=1.8\columnwidth]{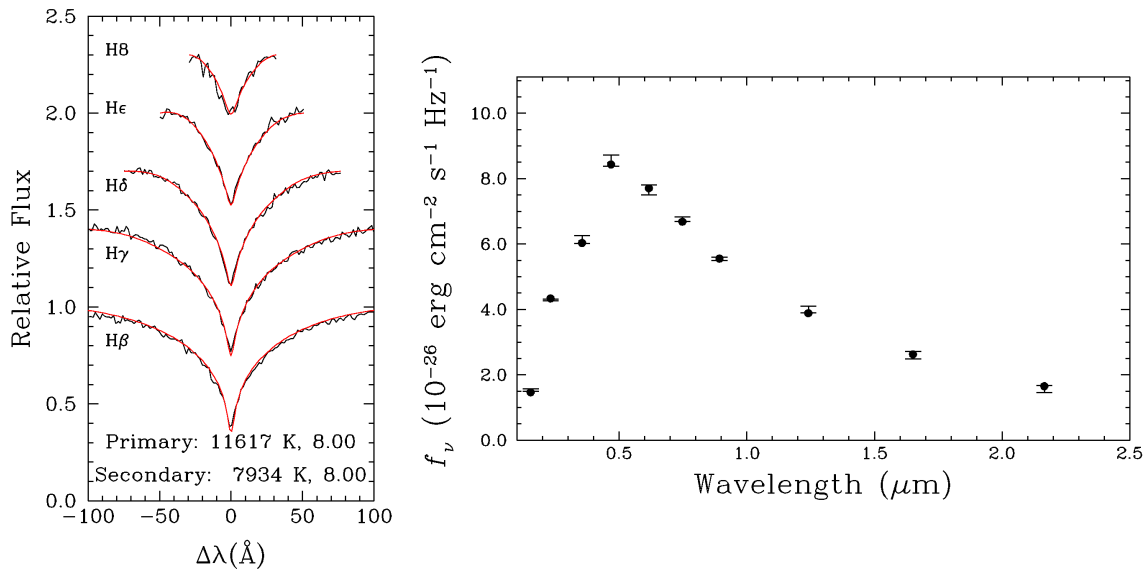}\caption{\label{doubleWD}  Simultaneous two white dwarf best model fit to the Balmer lines (left panel) and the $fuv,nuv,u,g,r,i,z,J,H,K$ photometry (right panel) of WD1310+583. The optical spectrum available does not cover the H$\alpha$ line.}
\end{figure*}

We attempted to fit both the optical spectrum and the FUV to NIR photometry of WD1310+583 simultaneously with two white dwarf models scaled to the same distance (Fig.\,\ref{doubleWD}). In both the spectroscopic and photometric fit the log\,$g$ values of the two stars are degenerate and cannot be realistically determined; so we fixed them at the canonical value of log\,$g=8.0$. We obtain a satisfactory fit with two white dwarfs with \Teff$=11\,617\pm70$ K and \Teff$=7934\pm290$ K (both values already include 3D corrections). This fit shows that a two white dwarf solution is indeed possible, but the \Teff values should be treated as purely indicative until Gaia parallax measurements can constrain both surface gravities.

\begin{figure}
\includegraphics[width=\columnwidth]{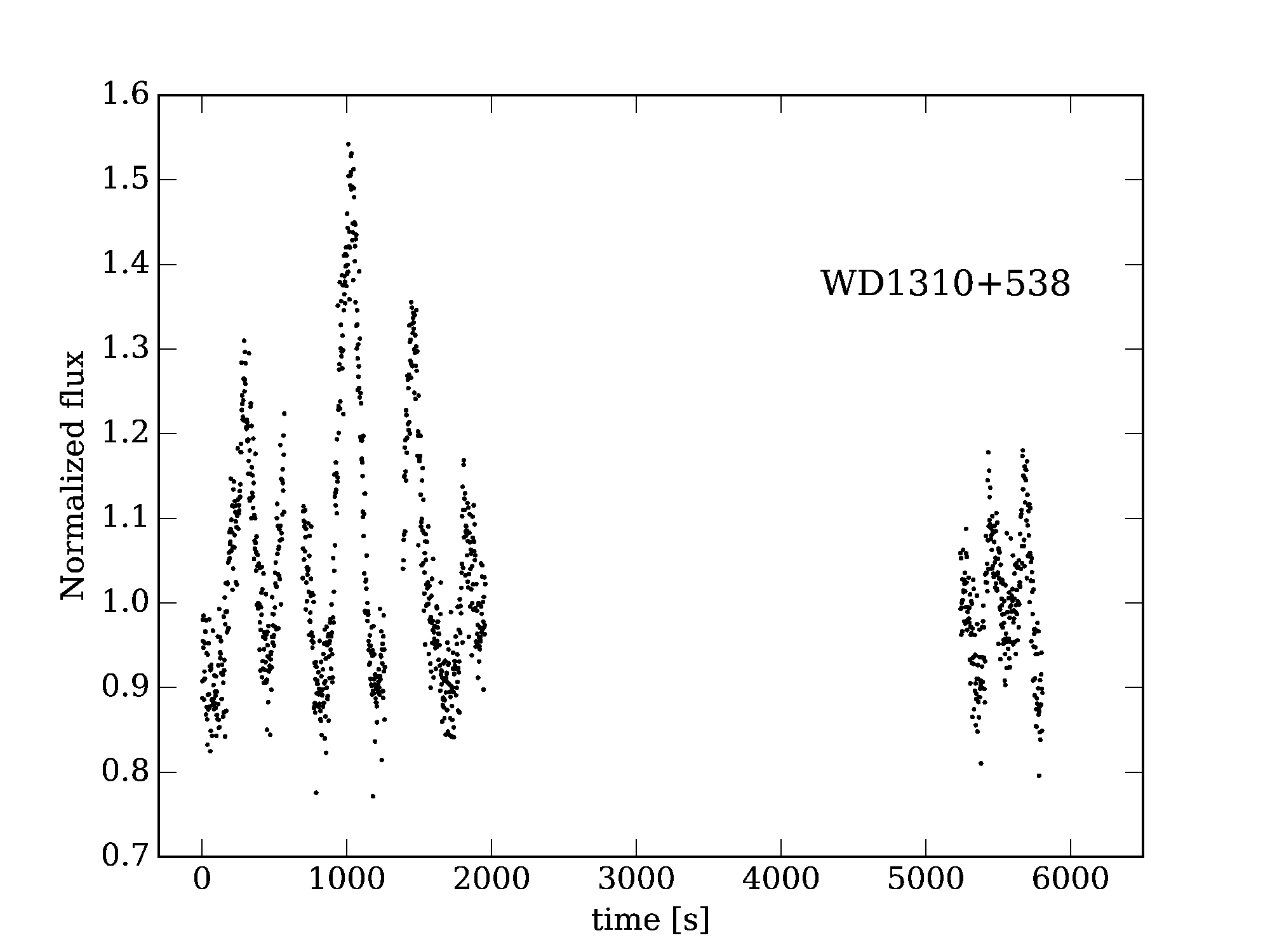}\caption{\label{lc_plot} Lightcurve of WD1310+538 obtained using the time-tag information in the COS spectrum. The lightcurve covers four 570 seconds observations using 2 seconds binning.}
\end{figure}

Interestingly, with log\,$g=8.0$ and \Teff = 11\,617\,K the hotter white dwarf would be inside the ZZ\,Ceti instability strip, while a white dwarf with \Teff$ = 10\,479$\,K (the solution from optical spectroscopy for a single white dwarf) would not. Indeed pulsations have been independently observed in this star (Zs. Bogn\'ar private communication). We used the time-tag information available with the COS spectrum to construct a lightcurve of WD1310+583 (Fig.\,\ref{lc_plot}) and our preliminary analysis reveals two potential pulsation periods of $390.88\pm1.32 s$ and $545.50\pm2.60 s$, consistent with ZZ\,Ceti pulsations. The detection of pulsation is proof of the presence of the hotter white dwarf and indirectly confirms the binary nature of this system. 

\section{Conclusions}
We analyzed optical spectroscopy and newly acquired COS FUV spectroscopy of the magnetic white dwarf WD2105$-$820 using convective atmospheric models and novel radiative ones. For comparison we carried out the same spectroscopic analysis on three additional non-magnetic white dwarfs which, like WD2105$-$820, are all cool enough to have developed convective atmospheres. 
For two of our comparison stars (WD0839$-$327 and WD1544$-$374) we find that convective model atmospheric fits produce consistent FUV and optical \Teff values. However for WD2105$-$820 consistent \Teff values could only be obtained using radiative models while the best fit to optical and FUV spectra using convective models are significantly discrepant ($>3\sigma$).
Additionally, model fit to the $griJHK$ photometry of WD2108$-$820 also give a \Teff value consistent with our radiative spectroscopic fits, but over 500\,K cooler than the convective fits.

Our third comparison white dwarf, WD1310+583, appears to have inconsistent FUV and optical \Teff values when either the convective or radiative models are used. We speculate that WD1310+583 is in fact a double white dwarf system in which a hotter object dominates the FUV emission.

We conclude that, unlike other white dwarfs with similar \Teff , WD2105$-$820 appears to have a radiative atmosphere. These observations directly confirm the theory that weak magnetic fields ($B\simeq50$ kG), like the one harbored by WD2105$-$820, inhibit convection in the outer atmosphere of white dwarfs \citep{tremblayetal15-1}. 

\section*{Acknowledgements}
We are thankful to John Landstreet for his quick and constructive review.
We thank A. Gianninas for making his observations available.
This project has received funding from the European Research Council (ERC) under the European Union’s Horizon 2020 research and innovation programme (grant agreements No 677706 - WD3D and No 320964 - WDTracer). This
work is based on observations made with the NASA/ESA Hubble Space Telescope, obtained at the Space Telescope Science Institute, which is operated by the Association of Universities for Research in Astronomy, Inc., under NASA contract NAS 5-26555. These observations are associated with programs \#14214 and \#14076.

This publication makes use of data products from the Two Micron All Sky Survey, which is a joint project of the University of Massachusetts and the Infrared Processing and Analysis Center/California Institute of Technology, funded by the National Aeronautics and Space Administration and the National Science Foundation.

Funding for the Sloan Digital Sky Survey IV has been provided by
the Alfred P. Sloan Foundation, the U.S. Department of Energy Office of
Science, and the Participating Institutions. SDSS-IV acknowledges
support and resources from the Center for High-Performance Computing at
the University of Utah. The SDSS web site is www.sdss.org.

SDSS-IV is managed by the Astrophysical Research Consortium for the 
Participating Institutions of the SDSS Collaboration including the 
Brazilian Participation Group, the Carnegie Institution for Science, 
Carnegie Mellon University, the Chilean Participation Group, the French Participation Group, Harvard-Smithsonian Center for Astrophysics, 
Instituto de Astrof\'isica de Canarias, The Johns Hopkins University, 
Kavli Institute for the Physics and Mathematics of the Universe (IPMU) / 
University of Tokyo, Lawrence Berkeley National Laboratory, 
Leibniz Institut f\"ur Astrophysik Potsdam (AIP),  
Max-Planck-Institut f\"ur Astronomie (MPIA Heidelberg), 
Max-Planck-Institut f\"ur Astrophysik (MPA Garching), 
Max-Planck-Institut f\"ur Extraterrestrische Physik (MPE), 
National Astronomical Observatories of China, New Mexico State University, 
New York University, University of Notre Dame, 
Observat\'ario Nacional / MCTI, The Ohio State University, 
Pennsylvania State University, Shanghai Astronomical Observatory, 
United Kingdom Participation Group,
Universidad Nacional Aut\'onoma de M\'exico, University of Arizona, 
University of Colorado Boulder, University of Oxford, University of Portsmouth, 
University of Utah, University of Virginia, University of Washington, University of Wisconsin, 
Vanderbilt University, and Yale University.

\bibliographystyle{mnras}

\end{document}